\documentclass[pre,aps,amsmath,amssymb,amsfonts,floatfix,superscriptaddress,showpacs,footinbib,twocolumn]{revtex4-1}

\usepackage{graphicx}
\usepackage{amsmath,amsfonts}
\usepackage{xcolor}
\usepackage{color}
\usepackage{mathtools}
\usepackage{pgfplots}
\usepackage{comment}
\usepackage{soul}
\usepackage[normalem]{ulem}

\newcommand{\up}{\ensuremath{\uparrow}}
\newcommand{\dn}{\ensuremath{\downarrow}}
\newcommand{\kv}{\ensuremath{\mathbf{k}}}

\newcommand{\qv}{\ensuremath{\mathbf{q}}}
\newcommand{\Qv}{\ensuremath{\mathbf{Q}}}

\newcommand{\sz}{\ensuremath{\text{sp}}}

\usepackage{tikz}
\usetikzlibrary{calc}
\usetikzlibrary{decorations.pathmorphing}
\usetikzlibrary{decorations.pathreplacing}
\usetikzlibrary{decorations.markings}
\usetikzlibrary{shapes.misc}
\usetikzlibrary{positioning}
\usetikzlibrary{arrows}
\usetikzlibrary{fadings}
\usetikzlibrary{shapes.callouts,tikzmark}
\tikzfading[name=fade out,
inner color=transparent!0,
outer color=transparent!100]
\tikzset{snake it/.style={decorate, decoration=snake}}
\usetikzlibrary{3d}
\usepackage{mathtools}
    \tikzset{
            partial ellipse/.style args={#1:#2:#3}{
                        insert path={+ (#1:#3) arc (#1:#2:#3)}
                            }
                        }

\usetikzlibrary{calc}
\tikzset{
            inertial frame/.style = {x={(-20:2cm)}, y={(-160:2cm)}, z={(90:2cm)}},
              local frame/.style = {shift={(local origin)}, x={(40:.7cm)}, y={(150:.7cm)}, z={(105:.7cm)}}
          }
\tikzset{middlearrow/.style={
                decoration={markings,
                            mark= at position 0.5 with {\arrow{#1}} ,
                                    },
                                            postaction={decorate}
                                                }
                                                }
\tikzset{cross/.style={cross out, draw, 
         minimum size=2*(#1-\pgflinewidth), 
                  inner sep=0pt, outer sep=0pt}}
                  
\definecolor{cblue}{HTML}{034694}

\def\presuper#1#2%
  {\mathop{}%
   \mathopen{\vphantom{#2}}^{#1}%
   \kern-0.5\scriptspace%
   #2}

\usepackage{hyperref}
\hypersetup{colorlinks=true,breaklinks,linkcolor=blue,urlcolor=blue,citecolor=blue}

\begin{document}
%
%
\def\umphys{%
    Department of Physics, University of Michigan,
    Ann Arbor, MI 48109, USA
}%
\def\tuwien{
Institute for Solid State Physics, TU Wien, 1040 Vienna, Austria
}

\author{Yang Yu}
\affiliation{\umphys}
\author{Sergei Iskakov}
\affiliation{\umphys}
\author{Emanuel Gull}
\affiliation{\umphys}
\author{Karsten Held}
\affiliation{\tuwien}
\author{Friedrich Krien}
\affiliation{\tuwien}

\title{Unambiguous Fluctuation Decomposition of the Self-Energy:\\ Pseudogap Physics beyond Spin Fluctuations}

\begin{abstract}
Correlated electron systems may give rise to multiple effective interactions whose combined impact on quasiparticle properties can be difficult to disentangle. We introduce an unambiguous decomposition of the electronic self-energy which allows us to quantify the contributions of various effective interactions simultaneously. We use this tool to revisit the hole-doped Hubbard model within the dynamical cluster approximation, where commonly spin fluctuations are considered to be the origin of the pseudogap. While our fluctuation decomposition confirms that spin fluctuations indeed suppress antinodal electronic spectral weight, we show that they alone cannot capture the pseudogap self-energy quantitatively. Nonlocal multiboson Feynman diagrams yield substantial contributions and are needed for a quantitative description of the pseudogap.
\end{abstract}

\maketitle

{\textit{Introduction.}---} Interacting electrons can be described as noninteracting ones modified by a self-energy which renormalizes the excitation energy, provides a finite lifetime, and sometimes even splits excitations into multiplets. The self-energy thus carries the imprint of the bare and effective interactions. Hence, from a suitable protocol for the analysis of the self-energy one may draw conclusions about these interactions. A paradigmatic example is the electron-phonon coupling, which famously gives rise to superconductivity in, e.g., lead. In this case, the celebrated Migdal-Eliashberg theory~\cite{Migdal58,Eliashberg60} directly connects the electron-phonon coupling to the two-particle vertex and the electronic self-energy. The representation of the self-energy through a phonon propagator explains the manifestation of the Debye energy in the photoemission spectrum of the normal and superconducting states of lead~\cite{Reinert03}. From the perspective of physical understanding, it is helpful to identify the boson mediating the effective interaction, which in this case is the phonon.

However, the electron-phonon problem is one of particular transparency, since the effective interaction arises from the weak coupling of two separate subsystems with different energy scales. A perturbative treatment is thus possible and allows for Migdal's approximation~\cite{Migdal58}, which neglects the renormalization of the electron-phonon coupling. In other cases, such as in unconventional superconductors, the effective interaction may itself arise from electronic correlations. Not surprisingly, the dominant effective interaction, if it exists, is often unknown and we frequently witness the competition of multiple effective interactions. While there may be a candidate for the dominant effective interaction, e.g., spin fluctuations (paramagnons) in the case of unconventional superconductivity~\cite{Scalapino12}, an unequivocal identification and a quantification of the contribution of this leading boson is hitherto not possible. Moreover, vertex corrections may play a role, leading to the challenging task of disentangling the impact of multiple effective interactions (or bosons) in the absence of Migdal's theorem~\cite{Migdal58}. 

\begin{figure*}
\begin{center}
  \begin{tikzpicture}
\node (image1) at (0,0){\includegraphics[width=1\textwidth]{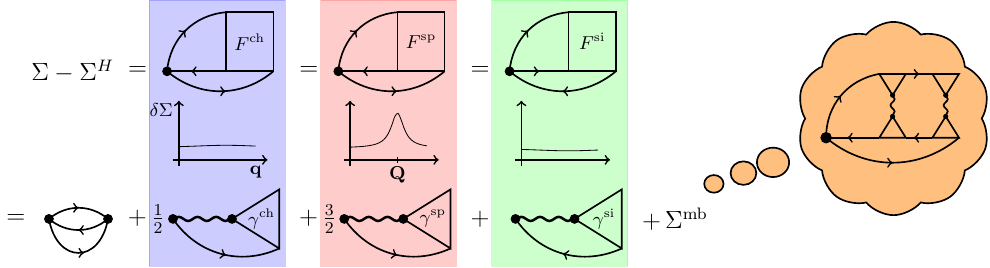}};
\end{tikzpicture}
\end{center}
\vspace{-0.7cm}
\caption{\label{fig:fdiag}
    Fluctuation diagnostics according to Ref.~\cite{Gunnarsson15} (top, center) and the fluctuation decomposition~\eqref{eq:fdiag} (bottom).
    Arrows, wiggly lines, triangles, and dots denote the Green's function, susceptibilities, Hedin vertices,
    and the Hubbard interaction, respectively.
    The top row shows the three equivalent diagrammatic representations ({\sl ``pictures''}) of the self-energy in terms of the charge (blue), spin (red), and singlet (green) four-point vertex. In the pseudogap regime the integrand is peaked around the momentum $\Qv=(\pi,\pi)$ in the spin picture, while it is flat in the charge and singlet pictures (center row diagrams). The right-hand-side of Eq.~\eqref{eq:fdiag}, shown at the bottom, allows us to evaluate the contribution of spin fluctuations quantitatively and unambiguously. An exemplary two-boson exchange diagram is shown on the right. For details, cf.~Supplemental Material~\cite{suppl}.}
\end{figure*}

A step in this direction is the method of fluctuation diagnostics in the Hubbard model~\cite{Gunnarsson15,Schaefer21-2}. In this approach, the electronic self-energy is represented in terms of the four-point vertex function $F$, which encapsulates the complete two-particle correlation information. Via the equation of motion, the self-energy is expressed in terms of $F$ as shown at the top of Fig.~\ref{fig:fdiag}. Here, arrows and filled circles represent the electronic Green's function and bare interaction, respectively. Because of a decoupling ambiguity of the Hubbard interaction $U\hat{n}_\uparrow\hat{n}_\downarrow$~\cite{Ayral17,Harkov21-2}, where $\hat{n}_\sigma$ is the density of electrons with spin $\sigma=\uparrow,\downarrow$, one can exactly represent the self-energy either in terms of the charge, spin, or singlet (particle-particle) vertex function $F^{\text{ch/sp/si}}$. By comparison of these pictures one may deduce whether the self-energy is  ``better'' described through charge, spin, or singlet fluctuations~\cite{Gunnarsson15,Schaefer21-2}. Namely, the self-energy is obtained from the vertex function $F(k,k',q)$ through summation over fermionic and bosonic momentum-frequency variables $k'=(\kv',\nu')$ and $q=(\qv,\omega)$, respectively. Then, the summation over the bosonic momentum may be peaked around a characteristic momentum $\Qv$, indicating a well-defined bosonic mode in the given picture, see the center row of Fig.~\ref{fig:fdiag}.

This protocol is, however, based on the ambiguity of the Hubbard interaction and thus it does not reveal the diagrammatic origin of the observed self-energy unequivocally. Thus, it cannot guide the formulation of analytic approximations to the self-energy, such as the Migdal-Eliashberg theory mentioned above. More specifically, all two-particle correlations are conflated into $F$ and therefore the quantitative contribution of a given fluctuation (or boson) to the self-energy cannot be estimated. As a result, the role of vertex corrections, which may alter the momentum dependence of the self-energy qualitatively, is difficult to assess~\cite{Krien22}. Finally, the approach introduced in Ref.~\cite{Gunnarsson15} is restricted to Hubbard models and cannot be straightforwardly generalized to, for example, nonlocal interactions or the electron-phonon coupling.

A second approach, which could in principle guide analytic approximations and which is applicable beyond the Hubbard model, is the parquet decomposition of the self-energy~\cite{Gunnarsson16}. Here, the vertex is decomposed into one irreducible and three reducible classes of Feynman diagrams. Unfortunately, this approach breaks down at strong coupling, where some of these vertices develop mutually canceling divergences. It is therefore unsuitable for the purpose of fluctuation analysis in this regime, which is relevant, e.g., for unconventional superconductivity. Hence, the question of a quantitative fluctuation decomposition has hitherto remained an open problem.

In this Letter, we introduce an unambiguous decomposition of the self-energy which remedies these shortcomings. It is based on the recently introduced single-boson exchange (SBE) decomposition of the vertex function~\cite{Krien19-4}. As shown in Supplemental Material~\cite{suppl}, this allows us to recast the self-energy into the form shown at the bottom of Fig.~\ref{fig:fdiag}, where the self-energy is given as a sum of charge, spin, and singlet fluctuation exchange diagrams, as well as a remainder. Triangles denote the three-legged fermion-boson (Hedin) vertex including {\sl all} vertex corrections. This decomposition allows for an unequivocal identification of the relevant boson, with the remainder containing genuine multiboson (mb) contributions.

Although similar, this representation of the self-energy is different from the Hedin equation~\cite{Hedin65}, which does not resolve the decoupling ambiguity of the Hubbard interaction~\cite{Ayral17}. Instead, the decomposition of the self-energy is unambiguous, that is, it can be derived from the equation of motion in any of the three pictures defining the fluctuation diagnostics approach (see Supplemental Material~\cite{suppl}). However, special care has to be taken of the diagram arising at the second order in the interaction $U$ {(see Fig.~\ref{fig:fdiag}, bottom left)}, which is recovered from the charge, spin, and singlet channels, respectively. Since these channels yield distinct Feynman diagrams only beyond $\mathcal{O}(U^2)$, we consider this diagram separately.

The self-energy decomposition defined below is exact and the various components can be calculated within exact or approximate frameworks. As an exemplary application, we consider here the hole-doped Hubbard model at strong coupling within the $8$-site dynamical cluster approximation (DCA,~\cite{Maier05}). For suitable temperatures and dopings the Hubbard model exhibits the so-called pseudogap, a suppression of antinodal electronic low-energy excitations, see, for example,  Refs.~\cite{Civelli05,Kyung06,Macridin06,Werner09,Gull09, Gull10,Sordi12,Gull13,Gunnarsson15, Wu17,Wu18,Krien22}. Investigations using fluctuation diagnostics~\cite{Gunnarsson15,Wu17,Wu22} showed that only the spin picture ($F^\text{sp}$) reveals a well-defined bosonic mode peaked around the momentum $\Qv=(\pi,\pi)$, as indicated by the center row of Fig.~\ref{fig:fdiag}. On closer inspection the bosonic Matsubara frequency $\omega=0$ dominates, indicating a thermal spin fluctuation~\cite{Gunnarsson15}. Here, we revisit the pseudogap self-energy with the approach depicted at the bottom of Fig.~\ref{fig:fdiag}, which allows for an unambiguous  quantitative estimate
of the various contributions. 
Our main finding is that while qualitatively spin fluctuations are responsible for opening the pseudogap, multiboson Feynman diagrams are of the same magnitude and partially compensate for these (single-boson) spin fluctuations.
Hence, an effective interaction mediated by spin fluctuations alone cannot describe the pseudogap quantitatively.

{\textit{Fluctuation decomposition.}---} We introduce the following decomposition of the self-energy,
\begin{align}
\Sigma_k-\Sigma^H=\Sigma^\text{2nd}_k+\Sigma^\text{ch}_k+\Sigma^\text{sp}_k+\Sigma^\text{si}_k+\Sigma^\text{mb}_k,\label{eq:fdiag}
\end{align}
where $\Sigma^H=U\langle\hat{n}_\sigma\rangle$ is the Hartree self-energy, $\Sigma^\text{2nd}_k=-U^2\sum_{qk'}G_{k+q}G_{k'}G_{k'+q}$ is the self-energy at second order in $U$, and $G_k$ denotes the electronic Green's function. Here and in the following, summations over momentum and frequencies imply a factor $T/N$, with $T$ being the temperature and $N$ the number of lattice (or cluster) sites. The individual contributions of single-boson exchange read as follows:
\begin{subequations}
\begin{align}
\Sigma^\text{ch}_k=&-\frac{U^2}{2}\sum_q G_{k+q}\gamma^\text{ch}_{kq}\chi^\text{ch}_q-\frac{1}{2}\Sigma^\text{2nd}_k,\label{eq:sbe_ch}\\
\Sigma^\text{sp}_k=&-\frac{3U^2}{2}\sum_q G_{k+q}\gamma^\text{sp}_{kq}\chi^\text{sp}_q-\frac{3}{2}\Sigma^\text{2nd}_k,\label{eq:sbe_sp}\\
\Sigma^\text{si}_k=&-U^2\sum_q G_{q-k}\gamma^\text{si}_{kq}\chi^\text{si}_q-\Sigma^\text{2nd}_k.\label{eq:sbe_si}
\end{align}
\end{subequations}
Here, $\chi_q$ denotes the susceptibility and $\gamma_{kq}$ the Hedin vertex~\cite{suppl}. The numerically exact calculation of the quantities in Eqs.~\eqref{eq:fdiag}-\eqref{eq:sbe_si} within DCA calculations is described in Supplemental Material~\cite{suppl}; without loss of generality, here $\kv$ and $\qv$ correspond to cluster momenta.

Note the similarity of Eqs.~\eqref{eq:sbe_ch}-\eqref{eq:sbe_si} to the fluctuation exchange (FLEX) approximation~\cite{Bickers89}, where $\gamma^\text{ch}, \gamma^\text{sp}, \gamma^\text{si}$ are set to their noninteracting values $1,1, \text{and} -1$. The unique prefactors $\frac{1}{2}, \frac{3}{2}, 1$ for the charge, spin, and singlet channels in the exact expressions~\eqref{eq:sbe_ch}-\eqref{eq:sbe_si} also arise naturally in crossing-symmetric approximations to the self-energy, such as, for example, FLEX or in the dynamical vertex approximation (D$\Gamma$A,~\cite{Toschi07}).

Considering the noninteracting limits of $\chi$ and $\gamma$ it is easy to see that $\mathcal{O}(U^2)$ contributions cancel from Eqs.~\eqref{eq:sbe_ch}-\eqref{eq:sbe_si}. The quantities $\Sigma^{\text{sp/ch/si}}$ are analogous to the three {\it pictures} of the fluctuation diagnostics~\cite{Gunnarsson15} where a charge, spin, or singlet fluctuation, peaked at a characteristic momentum $\Qv$, coincides with a peak in the respective integrand (cf. Fig.~\ref{fig:fdiag}). However, in striking contrast to the mutually exclusive choice of a picture within fluctuation diagnostics, these terms contribute to the full self-energy in the unambiguous decomposition~\eqref{eq:fdiag} through their sum. 

Finally, $\Sigma^\text{mb}$ includes Feynman diagrams that cannot be cast in terms of the exchange of a single boson; it is fundamentally of a {\sl multi}-bosonic character and corresponding Feynman diagrams include at least two or more fluctuations, as indicated on the right of Fig.~\ref{fig:fdiag}. It is not possible to bring these diagrams into the form of Eqs.~\eqref{eq:sbe_ch}-\eqref{eq:sbe_si}. Therefore, $\Sigma^\text{mb}$ represents an independent class of Feynman diagrams. We show in the following that its contribution is sizable in the hole-doped Hubbard model at strong coupling and that it differs qualitatively from $\Sigma^\text{sp}$, which captures the spin fluctuations. This may explain the difficulty in understanding the strongly correlated regime of the Hubbard model in terms of simple analytic approximations.

\begin{figure}[b]
\begin{center}
  \begin{tikzpicture}
\node (image1) at (0,0){\includegraphics[width=0.48\textwidth]{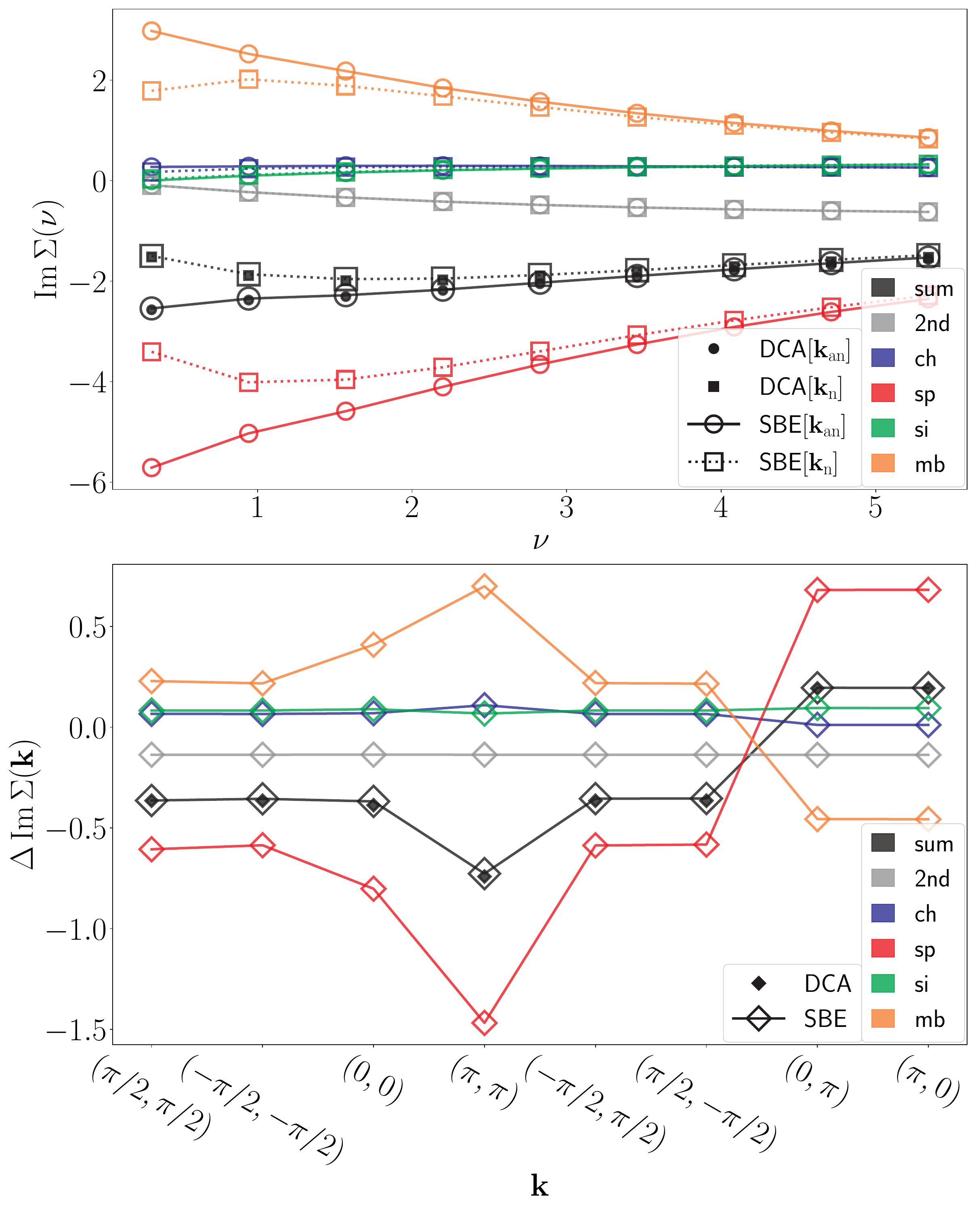}};
\end{tikzpicture}
\end{center}
\vspace{-0.7cm}
\caption{Top: 8-site DCA self-energy in the pseudogap regime at the nodal (squares) and antinodal (circles) points. The full self-energy is shown in black; the decomposition into channels in color; and the second order term in gray. Bottom: Slope of the self-energy
at small Matsubara frequencies along a high-symmetry path through the Brillouin zone. Physically, this slope yields an estimate for the enhancement of the effective mass $m^*/m= 1-\Delta\text{Im}\Sigma({\mathbf k})/(2\pi T)$ {\it if negative}, while positive slopes indicate insulating behavior.
Parameters: $\langle\hat{n}\rangle=0.95$, $U=7t$, $t'=-0.15t$, $T=0.1t$.
}
\label{fig:pg}
\end{figure}

{\textit{Results.}---}As an application, we analyze, in light of the unambiguous decomposition~\eqref{eq:fdiag}, the hole-doped Hubbard model with parameters $t=1, U=7t, t'=-0.15t$ using the $8$-site DCA~\cite{Maier05}; the temperature is fixed to $T=0.1t$. The top panel of Fig.~\ref{fig:pg} shows the imaginary part of DCA self-energy at the nodal [$\kv_\text{n}=(\pi/2,\pi/2)$; filled black square] and antinodal point [$\kv_\text{an}=(\pi,0)$; filled black dot]. For the chosen density $\langle\hat{n}\rangle=\langle\hat{n}_\up\rangle+\langle\hat{n}_\dn\rangle=0.95$ and temperature, the difference between the self-energy at the second and first Matsubara frequencies, $\Delta\text{Im}\Sigma(\kv)=\text{Im}\Sigma(\kv,3\pi T)-\text{Im}\Sigma(\kv,\pi T)$ [see bottom panel of Fig.~\ref{fig:pg}], and thus the slope of the self-energy on Matsubara frequencies is negative (positive) at the node (antinode), indicating the dichotomy of the pseudogap regime~\cite{Schaefer15}, see also Fig. 9 of Ref.~\cite{Gull09}.

To verify the correctness of Eq.~\eqref{eq:fdiag}, the top panel of Fig.~\ref{fig:pg} also shows the sum of all the terms on its right-hand-side (open black symbols), which lies on top of the directly computed DCA self-energy (i.e., the left-hand-side), as it should. Next, we consider the various contributions to $\text{Im}\Sigma$ one by one. In the chosen strong-coupling regime the second-order self-energy $\text{Im}\Sigma^\text{2nd}$ (gray) is irrelevant for small Matsubara frequencies, however, it determines the $\frac{1}{\nu}$-asymptote of the self-energy~\cite{Kugler22}, as shown in Supplemental Material~\cite{suppl}.

\begin{figure}[t]
\begin{center}
  \begin{tikzpicture}
\node (image1) at (0,0){\includegraphics[width=0.48\textwidth]{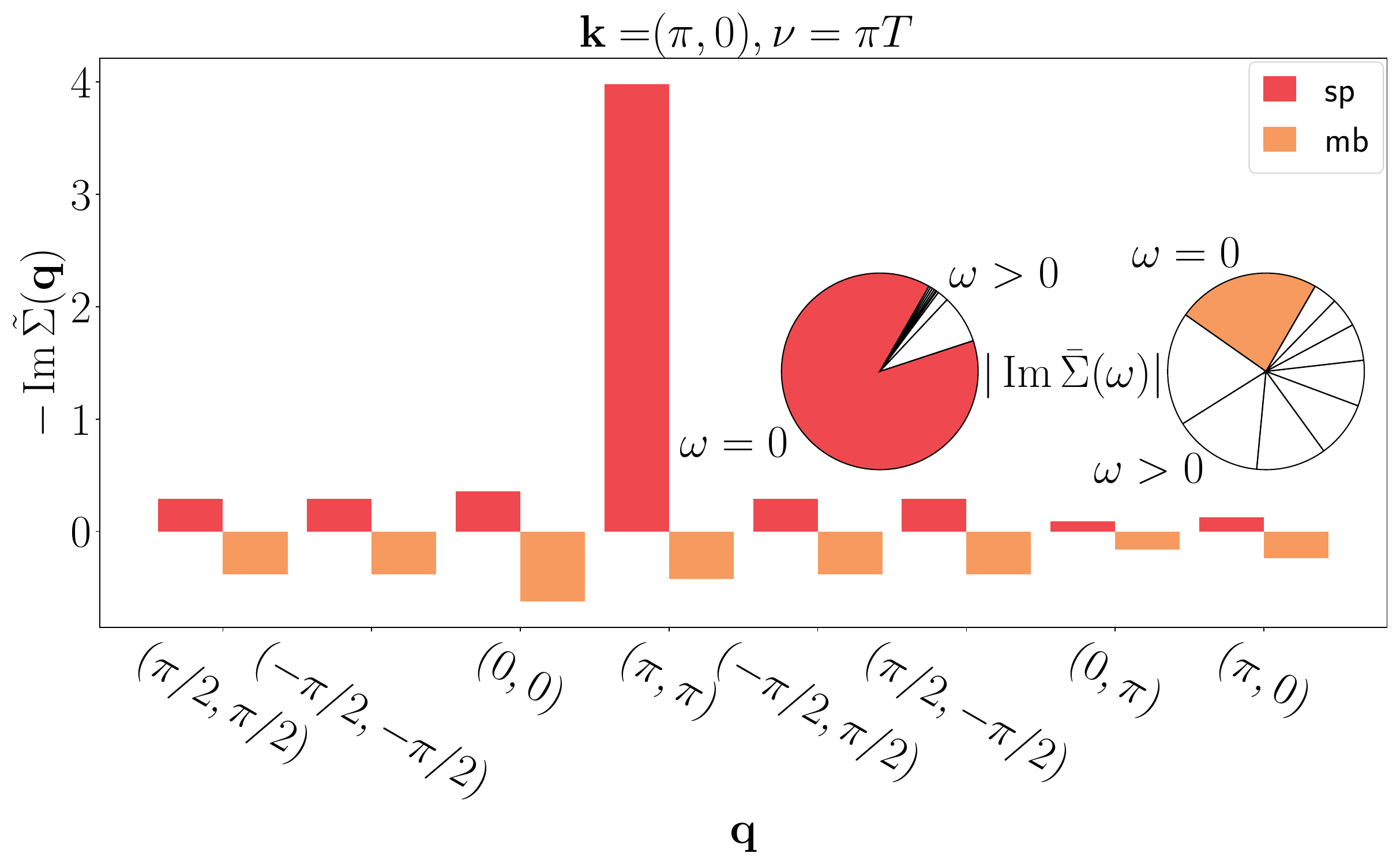}};
\end{tikzpicture}
\end{center}
\vspace{-0.7cm}
\caption{Bar chart: momentum($\qv$)-resolved fluctuation decomposition of $\Sigma^\sz$ and $\Sigma^\text{mb}$ for the antinodal fermionic momentum and frequency $\nu=\pi T$ in the pseudogap regime. Pie charts: frequency($\omega$)-resolved decomposition.}
\label{fig:fdiag_q}
\end{figure}

Likewise, the charge and singlet single-boson exchange contributions $\text{Im}\Sigma^\text{ch}$ (blue) and $\text{Im}\Sigma^\text{si}$ (green) are negligible for small Matsubara frequencies, and hence for the pseudogap in the full self-energy. These quantities have a finite $\frac{1}{\nu}$-asymptote, but $\text{Im}\Sigma^\text{ch}$ and $\text{Im}\Sigma^\text{si}$ cancel asymptotically with $\text{Im}\Sigma^\text{sp}$~\cite{suppl}. A major contribution at small Matsubara frequencies stems from single spin-fluctuation exchange. $\text{Im}\Sigma^\text{sp}$ (red) indeed exhibits the nodal-antinodal dichotomy, similar to the full self-energy, but its absolute value is much bigger. Remarkably, $\text{Im}\Sigma^\text{mb}$ (orange) is of similar magnitude but has the opposite sign compared to $\text{Im}\Sigma^\text{sp}$.

We therefore arrive at the following twofold result: (i) Spin fluctuations open the pseudogap through their contribution $\Sigma^\text{sp}$ to the self-energy, as is clear from inspection of the bottom panel of Fig.~\ref{fig:pg}, which yields the only significant positive contribution to $\Delta\text{Im}\Sigma$ at the antinode. This is consistent with conclusions drawn in previous studies~\cite{Gunnarsson15,Wu17,Krien22}. (ii) Nevertheless, the self-energy cannot be described quantitatively through spin-fluctuation exchange alone, even if all Feynman diagrams for the spin-fermion vertex $\gamma^\text{sp}$ are taken into account, as we have done here. The multiboson exchange $\Sigma^\text{mb}$ is of similar magnitude as $\Sigma^\text{sp}$. The two quantities largely cancel each other out, but the full self-energy inherits the suppression of antinodal spectral weight from $\Sigma^\text{sp}$.

This cancellation may be related to similar effects observed in previous works. Within the parquet decomposition, a cancellation of the spin channel with various other diagrams was observed (see Fig. 8 of Ref.~\cite{Gunnarsson16}), whose origin could, however, not be determined due to the ill-behaved (divergent) behavior of this approach in the strong-coupling regime. Further, within fluctuation diagnostics, a cancellation was observed between contributions from different wave vectors $\qv$ in the spin picture, see Fig. 2 of Ref.~\cite{Gunnarsson15}, and a cancellation was also observed in the anomalous self-energy, see Fig. 3 of Ref.~\cite{Dong22-2}. Remarkably, the fluctuation decomposition~\eqref{eq:fdiag} implies that these cancellations actually occur between quantities of completely different origins, as we show in the following.

To this end, we consider further aspects of the important contribution $\Sigma^\text{mb}$. It is large also for larger doping~\cite{suppl} and thus its magnitude is not directly related to the pseudogap. It appears that a sizable $\Sigma^\text{mb}$ contribution arises directly from the large Hubbard interaction. Indeed, within the single-site dynamical mean-field theory (DMFT,~\cite{Georges96}) this quantity grows as $U$ approaches the Mott regime, see Supplemental Material~\cite{suppl}. On the other hand, we find in our DCA calculations that $\Sigma^\text{mb}$ exhibits a similar strong $\kv$ dependence as $\Sigma^\sz$ (cf. Fig.~\ref{fig:pg}, top), therefore, a local approximation would be insufficient.

The origin of $\Sigma^\text{mb}$ differs  qualitatively from $\Sigma^\sz$. To show this, we refine our fluctuation decomposition of the antinodal self-energy at $\nu=\pi T$, i.e., we resolve for the contributions of different bosonic momentum $\qv$ in the sum of Eq.~\eqref{eq:sbe_sp} (while the sum over $\omega$ is still performed).  The result is shown as the bar chart in Fig.~\ref{fig:fdiag_q}, denoted as $\tilde{\Sigma}(\qv)$. As anticipated by the red segment in Fig.~\ref{fig:fdiag}, the contributions to $\Sigma^\sz$ are indeed peaked at $\Qv=(\pi,\pi)$. Omitting the summation over $\omega$ (instead of $\qv$) results in the frequency-resolved contributions $\bar{\Sigma}(\omega)$. The red (left) pie chart in Fig.~\ref{fig:fdiag_q} confirms the dominance of $\omega=0$ in the spin channel.

On the other hand, a refined analysis of $\Sigma^\text{mb}$ reveals a different picture. In fact, although the decomposition in Eq.~\eqref{eq:fdiag} identifies the contribution of single charge, spin, and singlet fluctuations, as well as the multiboson exchange unambiguously, a residual ambiguity remains in the representation of $\Sigma^\text{mb}$ through Feynman diagrams, see Supplemental Material~\cite{suppl}. This means that it can be analyzed again in three different pictures in the sense of Ref.~\cite{Gunnarsson15}. We choose the spin picture and the momentum-resolved fluctuation diagnostics of $\Sigma^\text{mb}$ is shown in orange in Fig.~\ref{fig:fdiag_q}. It is not peaked at a particular momentum; similarly, the orange pie chart on the right shows that $\omega=0$ is not dominant either, both in stark contrast to $\Sigma^\sz$. We conclude that the distributed contributions to $\Sigma^\text{mb}$ were overshadowed {by $\Sigma^\sz$} in previous analyses~\cite{Gunnarsson15,Wu17}. Choosing a different picture for $\Sigma^\text{mb}$ does not change this result~\cite{suppl}. We do not observe divergent or otherwise ill-conditioned behavior of the remainder $\Sigma^\text{mb}$ in this or in any other investigated regime, overcoming the intrinsic difficulties of the approach presented in Ref.~\cite{Gunnarsson16}.

{\textit{Conclusions.}---}
We introduced an unambiguous decomposition of the self-energy, which allows us to identify the dominant boson and even to quantify its contributions to the self-energy. As an important application, we investigated the quantitative contribution of spin fluctuations to the self-energy of the hole-doped Hubbard model within DCA. Our analysis confirms that spin fluctuations open the pseudogap in the underdoped regime~\cite{Civelli05,Kyung06,Macridin06,Werner09,Gull09, Gull10,Sordi12,Gull13,Gunnarsson15, Wu17,Wu18,Krien22}. In addition, however, {we observe a second major source of nonlocal correlations which} cannot be associated with conventional spin fluctuations or any other bosonic mode. It is instead a multiboson excitation without the dominance of a thermal ($\omega=0$) frequency nor an antiferromagnetic [$\Qv=(\pi,\pi)$]  momentum which are characteristic of spin fluctuations. This shows that the picture of a spin-fluctuation-mediated pseudogap, while capturing the dominant contribution, is incomplete.

It is an intriguing outlook to formulate the fluctuation decomposition also for the {\it d}-wave ordered phase~\cite{Dong22,Dong22-2}. A generalization of the presented approach to multiple orbitals or nonlocal interactions is straightforward.

We thank Alessandro Toschi for useful comments. Y.Y. thanks Xinyang Dong for the helpful discussions regarding DCA. Y.Y. and E.G. were supported by the National Science Foundation under Grant No. NSF DMR 2001465; F.K. and K.H. by the Austrian Science Fund (FWF) projects  P32044,   P36213, V1018, 
 SFB Q-M\&S (FWF project ID F86), and Research Unit QUAST by the Deutsche Foschungsgemeinschaft 
 (DFG; project ID FOR5249) and FWF (project ID I 5868). This work used Expanse at SDSC through allocation DMR130036 from the Advanced Cyberinfrastructure Coordination Ecosystem: Services \& Support (ACCESS) program~\cite{access}, which is supported by National Science Foundation Grants No. 2138259, No. 2138286, No. 2138307,  No. 2137603, and No. 2138296.

\bibliography{main}

\begin{thebibliography}{40}%
\makeatletter
\providecommand \@ifxundefined [1]{%
 \@ifx{#1\undefined}
}%
\providecommand \@ifnum [1]{%
 \ifnum #1\expandafter \@firstoftwo
 \else \expandafter \@secondoftwo
 \fi
}%
\providecommand \@ifx [1]{%
 \ifx #1\expandafter \@firstoftwo
 \else \expandafter \@secondoftwo
 \fi
}%
\providecommand \natexlab [1]{#1}%
\providecommand \enquote  [1]{``#1''}%
\providecommand \bibnamefont  [1]{#1}%
\providecommand \bibfnamefont [1]{#1}%
\providecommand \citenamefont [1]{#1}%
\providecommand \href@noop [0]{\@secondoftwo}%
\providecommand \href [0]{\begingroup \@sanitize@url \@href}%
\providecommand \@href[1]{\@@startlink{#1}\@@href}%
\providecommand \@@href[1]{\endgroup#1\@@endlink}%
\providecommand \@sanitize@url [0]{\catcode `\\12\catcode `\$12\catcode `\&12\catcode `\#12\catcode `\^12\catcode `\_12\catcode `\%12\relax}%
\providecommand \@@startlink[1]{}%
\providecommand \@@endlink[0]{}%
\providecommand \url  [0]{\begingroup\@sanitize@url \@url }%
\providecommand \@url [1]{\endgroup\@href {#1}{\urlprefix }}%
\providecommand \urlprefix  [0]{URL }%
\providecommand \Eprint [0]{\href }%
\providecommand \doibase [0]{http://dx.doi.org/}%
\providecommand \selectlanguage [0]{\@gobble}%
\providecommand \bibinfo  [0]{\@secondoftwo}%
\providecommand \bibfield  [0]{\@secondoftwo}%
\providecommand \translation [1]{[#1]}%
\providecommand \BibitemOpen [0]{}%
\providecommand \bibitemStop [0]{}%
\providecommand \bibitemNoStop [0]{.\EOS\space}%
\providecommand \EOS [0]{\spacefactor3000\relax}%
\providecommand \BibitemShut  [1]{\csname bibitem#1\endcsname}%
\let\auto@bib@innerbib\@empty
\bibitem [{\citenamefont {Migdal}(1958)}]{Migdal58}%
  \BibitemOpen
  \bibfield  {author} {\bibinfo {author} {\bibfnamefont {A.~B.}\ \bibnamefont {Migdal}},\ }\href@noop {} {\bibfield  {journal} {\bibinfo  {journal} {Sov. Phys. JETP}\ }\textbf {\bibinfo {volume} {34}},\ \bibinfo {pages} {996} (\bibinfo {year} {1958})}\BibitemShut {NoStop}%
\bibitem [{\citenamefont {Eliashberg}(1960)}]{Eliashberg60}%
  \BibitemOpen
  \bibfield  {author} {\bibinfo {author} {\bibfnamefont {G.~M.}\ \bibnamefont {Eliashberg}},\ }\href@noop {} {\bibfield  {journal} {\bibinfo  {journal} {Sov. Phys. JETP}\ }\textbf {\bibinfo {volume} {11}},\ \bibinfo {pages} {696} (\bibinfo {year} {1960})}\BibitemShut {NoStop}%
\bibitem [{\citenamefont {Reinert}\ \emph {et~al.}(2003)\citenamefont {Reinert}, \citenamefont {Eltner}, \citenamefont {Nicolay}, \citenamefont {Ehm}, \citenamefont {Schmidt},\ and\ \citenamefont {H\"ufner}}]{Reinert03}%
  \BibitemOpen
  \bibfield  {author} {\bibinfo {author} {\bibfnamefont {F.}~\bibnamefont {Reinert}}, \bibinfo {author} {\bibfnamefont {B.}~\bibnamefont {Eltner}}, \bibinfo {author} {\bibfnamefont {G.}~\bibnamefont {Nicolay}}, \bibinfo {author} {\bibfnamefont {D.}~\bibnamefont {Ehm}}, \bibinfo {author} {\bibfnamefont {S.}~\bibnamefont {Schmidt}}, \ and\ \bibinfo {author} {\bibfnamefont {S.}~\bibnamefont {H\"ufner}},\ }\href {\doibase 10.1103/PhysRevLett.91.186406} {\bibfield  {journal} {\bibinfo  {journal} {Phys. Rev. Lett.}\ }\textbf {\bibinfo {volume} {91}},\ \bibinfo {pages} {186406} (\bibinfo {year} {2003})}\BibitemShut {NoStop}%
\bibitem [{\citenamefont {Scalapino}(2012)}]{Scalapino12}%
  \BibitemOpen
  \bibfield  {author} {\bibinfo {author} {\bibfnamefont {D.~J.}\ \bibnamefont {Scalapino}},\ }\href {\doibase 10.1103/RevModPhys.84.1383} {\bibfield  {journal} {\bibinfo  {journal} {Rev. Mod. Phys.}\ }\textbf {\bibinfo {volume} {84}},\ \bibinfo {pages} {1383} (\bibinfo {year} {2012})}\BibitemShut {NoStop}%
\bibitem [{\citenamefont {Gunnarsson}\ \emph {et~al.}(2015)\citenamefont {Gunnarsson}, \citenamefont {Sch\"afer}, \citenamefont {LeBlanc}, \citenamefont {Gull}, \citenamefont {Merino}, \citenamefont {Sangiovanni}, \citenamefont {Rohringer},\ and\ \citenamefont {Toschi}}]{Gunnarsson15}%
  \BibitemOpen
  \bibfield  {author} {\bibinfo {author} {\bibfnamefont {O.}~\bibnamefont {Gunnarsson}}, \bibinfo {author} {\bibfnamefont {T.}~\bibnamefont {Sch\"afer}}, \bibinfo {author} {\bibfnamefont {J.~P.~F.}\ \bibnamefont {LeBlanc}}, \bibinfo {author} {\bibfnamefont {E.}~\bibnamefont {Gull}}, \bibinfo {author} {\bibfnamefont {J.}~\bibnamefont {Merino}}, \bibinfo {author} {\bibfnamefont {G.}~\bibnamefont {Sangiovanni}}, \bibinfo {author} {\bibfnamefont {G.}~\bibnamefont {Rohringer}}, \ and\ \bibinfo {author} {\bibfnamefont {A.}~\bibnamefont {Toschi}},\ }\href {\doibase 10.1103/PhysRevLett.114.236402} {\bibfield  {journal} {\bibinfo  {journal} {Phys. Rev. Lett.}\ }\textbf {\bibinfo {volume} {114}},\ \bibinfo {pages} {236402} (\bibinfo {year} {2015})}\BibitemShut {NoStop}%
\bibitem [{\citenamefont {Schäfer}\ and\ \citenamefont {Toschi}(2021)}]{Schaefer21-2}%
  \BibitemOpen
  \bibfield  {author} {\bibinfo {author} {\bibfnamefont {T.}~\bibnamefont {Schäfer}}\ and\ \bibinfo {author} {\bibfnamefont {A.}~\bibnamefont {Toschi}},\ }\href {\doibase 10.1088/1361-648x/abeb44} {\bibfield  {journal} {\bibinfo  {journal} {Journal of Physics: Condensed Matter}\ }\textbf {\bibinfo {volume} {33}},\ \bibinfo {pages} {214001} (\bibinfo {year} {2021})}\BibitemShut {NoStop}%
\bibitem [{\citenamefont {Ayral}\ \emph {et~al.}(2017)\citenamefont {Ayral}, \citenamefont {Vu\ifmmode \check{c}\else \v{c}\fi{}i\ifmmode \check{c}\else \v{c}\fi{}evi\ifmmode~\acute{c}\else \'{c}\fi{}},\ and\ \citenamefont {Parcollet}}]{Ayral17}%
  \BibitemOpen
  \bibfield  {author} {\bibinfo {author} {\bibfnamefont {T.}~\bibnamefont {Ayral}}, \bibinfo {author} {\bibfnamefont {J.}~\bibnamefont {Vu\ifmmode \check{c}\else \v{c}\fi{}i\ifmmode \check{c}\else \v{c}\fi{}evi\ifmmode~\acute{c}\else \'{c}\fi{}}}, \ and\ \bibinfo {author} {\bibfnamefont {O.}~\bibnamefont {Parcollet}},\ }\href {\doibase 10.1103/PhysRevLett.119.166401} {\bibfield  {journal} {\bibinfo  {journal} {Phys. Rev. Lett.}\ }\textbf {\bibinfo {volume} {119}},\ \bibinfo {pages} {166401} (\bibinfo {year} {2017})}\BibitemShut {NoStop}%
\bibitem [{\citenamefont {Harkov}\ \emph {et~al.}(2021{\natexlab{a}})\citenamefont {Harkov}, \citenamefont {Vandelli}, \citenamefont {Brener}, \citenamefont {Lichtenstein},\ and\ \citenamefont {Stepanov}}]{Harkov21-2}%
  \BibitemOpen
  \bibfield  {author} {\bibinfo {author} {\bibfnamefont {V.}~\bibnamefont {Harkov}}, \bibinfo {author} {\bibfnamefont {M.}~\bibnamefont {Vandelli}}, \bibinfo {author} {\bibfnamefont {S.}~\bibnamefont {Brener}}, \bibinfo {author} {\bibfnamefont {A.~I.}\ \bibnamefont {Lichtenstein}}, \ and\ \bibinfo {author} {\bibfnamefont {E.~A.}\ \bibnamefont {Stepanov}},\ }\href {\doibase 10.1103/PhysRevB.103.245123} {\bibfield  {journal} {\bibinfo  {journal} {Phys. Rev. B}\ }\textbf {\bibinfo {volume} {103}},\ \bibinfo {pages} {245123} (\bibinfo {year} {2021}{\natexlab{a}})}\BibitemShut {NoStop}%
\bibitem [{sup()}]{suppl}%
  \BibitemOpen
  \href@noop {} {}\bibinfo {howpublished} {See the Supplemental Material at [URL], which included Refs.~\cite{Krien20-2,rubtsov05,werner06,gull08,gull11,Kitatani19,Harkov21}, for the derivation of the formulas presented in the main text, along with supplementary numerical results.}\BibitemShut {Stop}%
\bibitem [{\citenamefont {Krien}\ \emph {et~al.}(2022)\citenamefont {Krien}, \citenamefont {Worm}, \citenamefont {Chalupa-Gantner}, \citenamefont {Toschi},\ and\ \citenamefont {Held}}]{Krien22}%
  \BibitemOpen
  \bibfield  {author} {\bibinfo {author} {\bibfnamefont {F.}~\bibnamefont {Krien}}, \bibinfo {author} {\bibfnamefont {P.}~\bibnamefont {Worm}}, \bibinfo {author} {\bibfnamefont {P.}~\bibnamefont {Chalupa-Gantner}}, \bibinfo {author} {\bibfnamefont {A.}~\bibnamefont {Toschi}}, \ and\ \bibinfo {author} {\bibfnamefont {K.}~\bibnamefont {Held}},\ }\href {\doibase 10.1038/s42005-022-01117-5} {\bibfield  {journal} {\bibinfo  {journal} {Communications Physics}\ }\textbf {\bibinfo {volume} {5}} (\bibinfo {year} {2022}),\ 10.1038/s42005-022-01117-5}\BibitemShut {NoStop}%
\bibitem [{\citenamefont {Gunnarsson}\ \emph {et~al.}(2016)\citenamefont {Gunnarsson}, \citenamefont {Sch\"afer}, \citenamefont {LeBlanc}, \citenamefont {Merino}, \citenamefont {Sangiovanni}, \citenamefont {Rohringer},\ and\ \citenamefont {Toschi}}]{Gunnarsson16}%
  \BibitemOpen
  \bibfield  {author} {\bibinfo {author} {\bibfnamefont {O.}~\bibnamefont {Gunnarsson}}, \bibinfo {author} {\bibfnamefont {T.}~\bibnamefont {Sch\"afer}}, \bibinfo {author} {\bibfnamefont {J.~P.~F.}\ \bibnamefont {LeBlanc}}, \bibinfo {author} {\bibfnamefont {J.}~\bibnamefont {Merino}}, \bibinfo {author} {\bibfnamefont {G.}~\bibnamefont {Sangiovanni}}, \bibinfo {author} {\bibfnamefont {G.}~\bibnamefont {Rohringer}}, \ and\ \bibinfo {author} {\bibfnamefont {A.}~\bibnamefont {Toschi}},\ }\href {\doibase 10.1103/PhysRevB.93.245102} {\bibfield  {journal} {\bibinfo  {journal} {Phys. Rev. B}\ }\textbf {\bibinfo {volume} {93}},\ \bibinfo {pages} {245102} (\bibinfo {year} {2016})}\BibitemShut {NoStop}%
\bibitem [{\citenamefont {Krien}\ \emph {et~al.}(2019)\citenamefont {Krien}, \citenamefont {Valli},\ and\ \citenamefont {Capone}}]{Krien19-4}%
  \BibitemOpen
  \bibfield  {author} {\bibinfo {author} {\bibfnamefont {F.}~\bibnamefont {Krien}}, \bibinfo {author} {\bibfnamefont {A.}~\bibnamefont {Valli}}, \ and\ \bibinfo {author} {\bibfnamefont {M.}~\bibnamefont {Capone}},\ }\href {\doibase 10.1103/PhysRevB.100.155149} {\bibfield  {journal} {\bibinfo  {journal} {Phys. Rev. B}\ }\textbf {\bibinfo {volume} {100}},\ \bibinfo {pages} {155149} (\bibinfo {year} {2019})}\BibitemShut {NoStop}%
\bibitem [{\citenamefont {Hedin}(1965)}]{Hedin65}%
  \BibitemOpen
  \bibfield  {author} {\bibinfo {author} {\bibfnamefont {L.}~\bibnamefont {Hedin}},\ }\href {\doibase 10.1103/PhysRev.139.A796} {\bibfield  {journal} {\bibinfo  {journal} {Phys. Rev.}\ }\textbf {\bibinfo {volume} {139}},\ \bibinfo {pages} {A796} (\bibinfo {year} {1965})}\BibitemShut {NoStop}%
\bibitem [{\citenamefont {Maier}\ \emph {et~al.}(2005)\citenamefont {Maier}, \citenamefont {Jarrell}, \citenamefont {Prushke},\ and\ \citenamefont {Hettler}}]{Maier05}%
  \BibitemOpen
  \bibfield  {author} {\bibinfo {author} {\bibfnamefont {T.~A.}\ \bibnamefont {Maier}}, \bibinfo {author} {\bibfnamefont {M.}~\bibnamefont {Jarrell}}, \bibinfo {author} {\bibfnamefont {T.}~\bibnamefont {Prushke}}, \ and\ \bibinfo {author} {\bibfnamefont {M.}~\bibnamefont {Hettler}},\ }\href {\doibase 10.1103/RevModPhys.77.1027} {\bibfield  {journal} {\bibinfo  {journal} {Rev. Mod. Phys.}\ }\textbf {\bibinfo {volume} {77}},\ \bibinfo {pages} {1027} (\bibinfo {year} {2005})}\BibitemShut {NoStop}%
\bibitem [{\citenamefont {Civelli}\ \emph {et~al.}(2005)\citenamefont {Civelli}, \citenamefont {Capone}, \citenamefont {Kancharla}, \citenamefont {Parcollet},\ and\ \citenamefont {Kotliar}}]{Civelli05}%
  \BibitemOpen
  \bibfield  {author} {\bibinfo {author} {\bibfnamefont {M.}~\bibnamefont {Civelli}}, \bibinfo {author} {\bibfnamefont {M.}~\bibnamefont {Capone}}, \bibinfo {author} {\bibfnamefont {S.~S.}\ \bibnamefont {Kancharla}}, \bibinfo {author} {\bibfnamefont {O.}~\bibnamefont {Parcollet}}, \ and\ \bibinfo {author} {\bibfnamefont {G.}~\bibnamefont {Kotliar}},\ }\href {\doibase 10.1103/PhysRevLett.95.106402} {\bibfield  {journal} {\bibinfo  {journal} {Phys. Rev. Lett.}\ }\textbf {\bibinfo {volume} {95}},\ \bibinfo {pages} {106402} (\bibinfo {year} {2005})}\BibitemShut {NoStop}%
\bibitem [{\citenamefont {Kyung}\ \emph {et~al.}(2006)\citenamefont {Kyung}, \citenamefont {Kancharla}, \citenamefont {S\'en\'echal}, \citenamefont {Tremblay}, \citenamefont {Civelli},\ and\ \citenamefont {Kotliar}}]{Kyung06}%
  \BibitemOpen
  \bibfield  {author} {\bibinfo {author} {\bibfnamefont {B.}~\bibnamefont {Kyung}}, \bibinfo {author} {\bibfnamefont {S.~S.}\ \bibnamefont {Kancharla}}, \bibinfo {author} {\bibfnamefont {D.}~\bibnamefont {S\'en\'echal}}, \bibinfo {author} {\bibfnamefont {A.-M.~S.}\ \bibnamefont {Tremblay}}, \bibinfo {author} {\bibfnamefont {M.}~\bibnamefont {Civelli}}, \ and\ \bibinfo {author} {\bibfnamefont {G.}~\bibnamefont {Kotliar}},\ }\href {\doibase 10.1103/PhysRevB.73.165114} {\bibfield  {journal} {\bibinfo  {journal} {Phys. Rev. B}\ }\textbf {\bibinfo {volume} {73}},\ \bibinfo {pages} {165114} (\bibinfo {year} {2006})}\BibitemShut {NoStop}%
\bibitem [{\citenamefont {Macridin}\ \emph {et~al.}(2006)\citenamefont {Macridin}, \citenamefont {Jarrell}, \citenamefont {Maier}, \citenamefont {Kent},\ and\ \citenamefont {D'Azevedo}}]{Macridin06}%
  \BibitemOpen
  \bibfield  {author} {\bibinfo {author} {\bibfnamefont {A.}~\bibnamefont {Macridin}}, \bibinfo {author} {\bibfnamefont {M.}~\bibnamefont {Jarrell}}, \bibinfo {author} {\bibfnamefont {T.}~\bibnamefont {Maier}}, \bibinfo {author} {\bibfnamefont {P.~R.~C.}\ \bibnamefont {Kent}}, \ and\ \bibinfo {author} {\bibfnamefont {E.}~\bibnamefont {D'Azevedo}},\ }\href {\doibase 10.1103/PhysRevLett.97.036401} {\bibfield  {journal} {\bibinfo  {journal} {Phys. Rev. Lett.}\ }\textbf {\bibinfo {volume} {97}},\ \bibinfo {pages} {036401} (\bibinfo {year} {2006})}\BibitemShut {NoStop}%
\bibitem [{\citenamefont {Werner}\ \emph {et~al.}(2009)\citenamefont {Werner}, \citenamefont {Gull}, \citenamefont {Parcollet},\ and\ \citenamefont {Millis}}]{Werner09}%
  \BibitemOpen
  \bibfield  {author} {\bibinfo {author} {\bibfnamefont {P.}~\bibnamefont {Werner}}, \bibinfo {author} {\bibfnamefont {E.}~\bibnamefont {Gull}}, \bibinfo {author} {\bibfnamefont {O.}~\bibnamefont {Parcollet}}, \ and\ \bibinfo {author} {\bibfnamefont {A.~J.}\ \bibnamefont {Millis}},\ }\href {\doibase 10.1103/PhysRevB.80.045120} {\bibfield  {journal} {\bibinfo  {journal} {Phys. Rev. B}\ }\textbf {\bibinfo {volume} {80}},\ \bibinfo {pages} {045120} (\bibinfo {year} {2009})}\BibitemShut {NoStop}%
\bibitem [{\citenamefont {Gull}\ \emph {et~al.}(2009)\citenamefont {Gull}, \citenamefont {Parcollet}, \citenamefont {Werner},\ and\ \citenamefont {Millis}}]{Gull09}%
  \BibitemOpen
  \bibfield  {author} {\bibinfo {author} {\bibfnamefont {E.}~\bibnamefont {Gull}}, \bibinfo {author} {\bibfnamefont {O.}~\bibnamefont {Parcollet}}, \bibinfo {author} {\bibfnamefont {P.}~\bibnamefont {Werner}}, \ and\ \bibinfo {author} {\bibfnamefont {A.~J.}\ \bibnamefont {Millis}},\ }\href {\doibase 10.1103/PhysRevB.80.245102} {\bibfield  {journal} {\bibinfo  {journal} {Phys. Rev. B}\ }\textbf {\bibinfo {volume} {80}},\ \bibinfo {pages} {245102} (\bibinfo {year} {2009})}\BibitemShut {NoStop}%
\bibitem [{\citenamefont {Gull}\ \emph {et~al.}(2010)\citenamefont {Gull}, \citenamefont {Ferrero}, \citenamefont {Parcollet}, \citenamefont {Georges},\ and\ \citenamefont {Millis}}]{Gull10}%
  \BibitemOpen
  \bibfield  {author} {\bibinfo {author} {\bibfnamefont {E.}~\bibnamefont {Gull}}, \bibinfo {author} {\bibfnamefont {M.}~\bibnamefont {Ferrero}}, \bibinfo {author} {\bibfnamefont {O.}~\bibnamefont {Parcollet}}, \bibinfo {author} {\bibfnamefont {A.}~\bibnamefont {Georges}}, \ and\ \bibinfo {author} {\bibfnamefont {A.~J.}\ \bibnamefont {Millis}},\ }\href {\doibase 10.1103/PhysRevB.82.155101} {\bibfield  {journal} {\bibinfo  {journal} {Phys. Rev. B}\ }\textbf {\bibinfo {volume} {82}},\ \bibinfo {pages} {155101} (\bibinfo {year} {2010})}\BibitemShut {NoStop}%
\bibitem [{\citenamefont {Sordi}\ \emph {et~al.}(2012)\citenamefont {Sordi}, \citenamefont {S\'emon}, \citenamefont {Haule},\ and\ \citenamefont {Tremblay}}]{Sordi12}%
  \BibitemOpen
  \bibfield  {author} {\bibinfo {author} {\bibfnamefont {G.}~\bibnamefont {Sordi}}, \bibinfo {author} {\bibfnamefont {P.}~\bibnamefont {S\'emon}}, \bibinfo {author} {\bibfnamefont {K.}~\bibnamefont {Haule}}, \ and\ \bibinfo {author} {\bibfnamefont {A.-M.~S.}\ \bibnamefont {Tremblay}},\ }\href {\doibase 10.1103/PhysRevLett.108.216401} {\bibfield  {journal} {\bibinfo  {journal} {Phys. Rev. Lett.}\ }\textbf {\bibinfo {volume} {108}},\ \bibinfo {pages} {216401} (\bibinfo {year} {2012})}\BibitemShut {NoStop}%
\bibitem [{\citenamefont {Gull}\ \emph {et~al.}(2013)\citenamefont {Gull}, \citenamefont {Parcollet},\ and\ \citenamefont {Millis}}]{Gull13}%
  \BibitemOpen
  \bibfield  {author} {\bibinfo {author} {\bibfnamefont {E.}~\bibnamefont {Gull}}, \bibinfo {author} {\bibfnamefont {O.}~\bibnamefont {Parcollet}}, \ and\ \bibinfo {author} {\bibfnamefont {A.~J.}\ \bibnamefont {Millis}},\ }\href {\doibase 10.1103/PhysRevLett.110.216405} {\bibfield  {journal} {\bibinfo  {journal} {Phys. Rev. Lett.}\ }\textbf {\bibinfo {volume} {110}},\ \bibinfo {pages} {216405} (\bibinfo {year} {2013})}\BibitemShut {NoStop}%
\bibitem [{\citenamefont {Wu}\ \emph {et~al.}(2017)\citenamefont {Wu}, \citenamefont {Ferrero}, \citenamefont {Georges},\ and\ \citenamefont {Kozik}}]{Wu17}%
  \BibitemOpen
  \bibfield  {author} {\bibinfo {author} {\bibfnamefont {W.}~\bibnamefont {Wu}}, \bibinfo {author} {\bibfnamefont {M.}~\bibnamefont {Ferrero}}, \bibinfo {author} {\bibfnamefont {A.}~\bibnamefont {Georges}}, \ and\ \bibinfo {author} {\bibfnamefont {E.}~\bibnamefont {Kozik}},\ }\href {\doibase 10.1103/PhysRevB.96.041105} {\bibfield  {journal} {\bibinfo  {journal} {Phys. Rev. B}\ }\textbf {\bibinfo {volume} {96}},\ \bibinfo {pages} {041105(R)} (\bibinfo {year} {2017})}\BibitemShut {NoStop}%
\bibitem [{\citenamefont {Wu}\ \emph {et~al.}(2018)\citenamefont {Wu}, \citenamefont {Scheurer}, \citenamefont {Chatterjee}, \citenamefont {Sachdev}, \citenamefont {Georges},\ and\ \citenamefont {Ferrero}}]{Wu18}%
  \BibitemOpen
  \bibfield  {author} {\bibinfo {author} {\bibfnamefont {W.}~\bibnamefont {Wu}}, \bibinfo {author} {\bibfnamefont {M.~S.}\ \bibnamefont {Scheurer}}, \bibinfo {author} {\bibfnamefont {S.}~\bibnamefont {Chatterjee}}, \bibinfo {author} {\bibfnamefont {S.}~\bibnamefont {Sachdev}}, \bibinfo {author} {\bibfnamefont {A.}~\bibnamefont {Georges}}, \ and\ \bibinfo {author} {\bibfnamefont {M.}~\bibnamefont {Ferrero}},\ }\href {\doibase 10.1103/PhysRevX.8.021048} {\bibfield  {journal} {\bibinfo  {journal} {Phys. Rev. X}\ }\textbf {\bibinfo {volume} {8}},\ \bibinfo {pages} {021048} (\bibinfo {year} {2018})}\BibitemShut {NoStop}%
\bibitem [{\citenamefont {Wú}\ \emph {et~al.}(2022)\citenamefont {Wú}, \citenamefont {Wang},\ and\ \citenamefont {Tremblay}}]{Wu22}%
  \BibitemOpen
  \bibfield  {author} {\bibinfo {author} {\bibfnamefont {W.}~\bibnamefont {Wú}}, \bibinfo {author} {\bibfnamefont {X.}~\bibnamefont {Wang}}, \ and\ \bibinfo {author} {\bibfnamefont {A.-M.}\ \bibnamefont {Tremblay}},\ }\href {\doibase 10.1073/pnas.2115819119} {\bibfield  {journal} {\bibinfo  {journal} {Proceedings of the National Academy of Sciences}\ }\textbf {\bibinfo {volume} {119}} (\bibinfo {year} {2022}),\ 10.1073/pnas.2115819119}\BibitemShut {NoStop}%
\bibitem [{\citenamefont {Bickers}\ and\ \citenamefont {Scalapino}(1989)}]{Bickers89}%
  \BibitemOpen
  \bibfield  {author} {\bibinfo {author} {\bibfnamefont {N.}~\bibnamefont {Bickers}}\ and\ \bibinfo {author} {\bibfnamefont {D.}~\bibnamefont {Scalapino}},\ }\href {\doibase https://doi.org/10.1016/0003-4916(89)90359-X} {\bibfield  {journal} {\bibinfo  {journal} {Annals of Physics}\ }\textbf {\bibinfo {volume} {193}},\ \bibinfo {pages} {206 } (\bibinfo {year} {1989})}\BibitemShut {NoStop}%
\bibitem [{\citenamefont {Toschi}\ \emph {et~al.}(2007)\citenamefont {Toschi}, \citenamefont {Katanin},\ and\ \citenamefont {Held}}]{Toschi07}%
  \BibitemOpen
  \bibfield  {author} {\bibinfo {author} {\bibfnamefont {A.}~\bibnamefont {Toschi}}, \bibinfo {author} {\bibfnamefont {A.~A.}\ \bibnamefont {Katanin}}, \ and\ \bibinfo {author} {\bibfnamefont {K.}~\bibnamefont {Held}},\ }\href {\doibase 10.1103/PhysRevB.75.045118} {\bibfield  {journal} {\bibinfo  {journal} {Phys. Rev. B}\ }\textbf {\bibinfo {volume} {75}},\ \bibinfo {pages} {045118} (\bibinfo {year} {2007})}\BibitemShut {NoStop}%
\bibitem [{\citenamefont {Sch\"afer}\ \emph {et~al.}(2015)\citenamefont {Sch\"afer}, \citenamefont {Geles}, \citenamefont {Rost}, \citenamefont {Rohringer}, \citenamefont {Arrigoni}, \citenamefont {Held}, \citenamefont {Bl\"umer}, \citenamefont {Aichhorn},\ and\ \citenamefont {Toschi}}]{Schaefer15}%
  \BibitemOpen
  \bibfield  {author} {\bibinfo {author} {\bibfnamefont {T.}~\bibnamefont {Sch\"afer}}, \bibinfo {author} {\bibfnamefont {F.}~\bibnamefont {Geles}}, \bibinfo {author} {\bibfnamefont {D.}~\bibnamefont {Rost}}, \bibinfo {author} {\bibfnamefont {G.}~\bibnamefont {Rohringer}}, \bibinfo {author} {\bibfnamefont {E.}~\bibnamefont {Arrigoni}}, \bibinfo {author} {\bibfnamefont {K.}~\bibnamefont {Held}}, \bibinfo {author} {\bibfnamefont {N.}~\bibnamefont {Bl\"umer}}, \bibinfo {author} {\bibfnamefont {M.}~\bibnamefont {Aichhorn}}, \ and\ \bibinfo {author} {\bibfnamefont {A.}~\bibnamefont {Toschi}},\ }\href {\doibase 10.1103/PhysRevB.91.125109} {\bibfield  {journal} {\bibinfo  {journal} {Phys. Rev. B}\ }\textbf {\bibinfo {volume} {91}},\ \bibinfo {pages} {125109} (\bibinfo {year} {2015})}\BibitemShut {NoStop}%
\bibitem [{\citenamefont {Chalupa-Gantner}\ \emph {et~al.}(2022)\citenamefont {Chalupa-Gantner}, \citenamefont {Kugler}, \citenamefont {Hille}, \citenamefont {von Delft}, \citenamefont {Andergassen},\ and\ \citenamefont {Toschi}}]{Kugler22}%
  \BibitemOpen
  \bibfield  {author} {\bibinfo {author} {\bibfnamefont {P.}~\bibnamefont {Chalupa-Gantner}}, \bibinfo {author} {\bibfnamefont {F.~B.}\ \bibnamefont {Kugler}}, \bibinfo {author} {\bibfnamefont {C.}~\bibnamefont {Hille}}, \bibinfo {author} {\bibfnamefont {J.}~\bibnamefont {von Delft}}, \bibinfo {author} {\bibfnamefont {S.}~\bibnamefont {Andergassen}}, \ and\ \bibinfo {author} {\bibfnamefont {A.}~\bibnamefont {Toschi}},\ }\href {\doibase 10.1103/PhysRevResearch.4.023050} {\bibfield  {journal} {\bibinfo  {journal} {Phys. Rev. Res.}\ }\textbf {\bibinfo {volume} {4}},\ \bibinfo {pages} {023050} (\bibinfo {year} {2022})}\BibitemShut {NoStop}%
\bibitem [{\citenamefont {Dong}\ \emph {et~al.}(2022{\natexlab{a}})\citenamefont {Dong}, \citenamefont {Re}, \citenamefont {Toschi},\ and\ \citenamefont {Gull}}]{Dong22-2}%
  \BibitemOpen
  \bibfield  {author} {\bibinfo {author} {\bibfnamefont {X.}~\bibnamefont {Dong}}, \bibinfo {author} {\bibfnamefont {L.~D.}\ \bibnamefont {Re}}, \bibinfo {author} {\bibfnamefont {A.}~\bibnamefont {Toschi}}, \ and\ \bibinfo {author} {\bibfnamefont {E.}~\bibnamefont {Gull}},\ }\href {\doibase 10.1073/pnas.2205048119} {\bibfield  {journal} {\bibinfo  {journal} {Proceedings of the National Academy of Sciences}\ }\textbf {\bibinfo {volume} {119}} (\bibinfo {year} {2022}{\natexlab{a}}),\ 10.1073/pnas.2205048119}\BibitemShut {NoStop}%
\bibitem [{\citenamefont {Georges}\ \emph {et~al.}(1996)\citenamefont {Georges}, \citenamefont {Kotliar}, \citenamefont {Krauth},\ and\ \citenamefont {Rozenberg}}]{Georges96}%
  \BibitemOpen
  \bibfield  {author} {\bibinfo {author} {\bibfnamefont {A.}~\bibnamefont {Georges}}, \bibinfo {author} {\bibfnamefont {G.}~\bibnamefont {Kotliar}}, \bibinfo {author} {\bibfnamefont {W.}~\bibnamefont {Krauth}}, \ and\ \bibinfo {author} {\bibfnamefont {M.~J.}\ \bibnamefont {Rozenberg}},\ }\href {\doibase 10.1103/RevModPhys.68.13} {\bibfield  {journal} {\bibinfo  {journal} {Rev. Mod. Phys.}\ }\textbf {\bibinfo {volume} {68}},\ \bibinfo {pages} {13} (\bibinfo {year} {1996})}\BibitemShut {NoStop}%
\bibitem [{\citenamefont {Dong}\ \emph {et~al.}(2022{\natexlab{b}})\citenamefont {Dong}, \citenamefont {Gull},\ and\ \citenamefont {Millis}}]{Dong22}%
  \BibitemOpen
  \bibfield  {author} {\bibinfo {author} {\bibfnamefont {X.}~\bibnamefont {Dong}}, \bibinfo {author} {\bibfnamefont {E.}~\bibnamefont {Gull}}, \ and\ \bibinfo {author} {\bibfnamefont {A.~J.}\ \bibnamefont {Millis}},\ }\href {\doibase 10.1038/s41567-022-01710-z} {\bibfield  {journal} {\bibinfo  {journal} {Nature Physics}\ }\textbf {\bibinfo {volume} {18}},\ \bibinfo {pages} {1293} (\bibinfo {year} {2022}{\natexlab{b}})}\BibitemShut {NoStop}%
\bibitem [{\citenamefont {Boerner}\ \emph {et~al.}(2023)\citenamefont {Boerner}, \citenamefont {Deems}, \citenamefont {Furlani}, \citenamefont {Knuth},\ and\ \citenamefont {Towns}}]{access}%
  \BibitemOpen
  \bibfield  {author} {\bibinfo {author} {\bibfnamefont {T.~J.}\ \bibnamefont {Boerner}}, \bibinfo {author} {\bibfnamefont {S.}~\bibnamefont {Deems}}, \bibinfo {author} {\bibfnamefont {T.~R.}\ \bibnamefont {Furlani}}, \bibinfo {author} {\bibfnamefont {S.~L.}\ \bibnamefont {Knuth}}, \ and\ \bibinfo {author} {\bibfnamefont {J.}~\bibnamefont {Towns}},\ }in\ \href {\doibase 10.1145/3569951.3597559} {\emph {\bibinfo {booktitle} {Practice and Experience in Advanced Research Computing}}},\ \bibinfo {series and number} {PEARC '23}\ (\bibinfo  {publisher} {Association for Computing Machinery},\ \bibinfo {address} {New York, NY, USA},\ \bibinfo {year} {2023})\ p.\ \bibinfo {pages} {173–176}\BibitemShut {NoStop}%
\bibitem [{\citenamefont {Krien}\ \emph {et~al.}(2021)\citenamefont {Krien}, \citenamefont {Kauch},\ and\ \citenamefont {Held}}]{Krien20-2}%
  \BibitemOpen
  \bibfield  {author} {\bibinfo {author} {\bibfnamefont {F.}~\bibnamefont {Krien}}, \bibinfo {author} {\bibfnamefont {A.}~\bibnamefont {Kauch}}, \ and\ \bibinfo {author} {\bibfnamefont {K.}~\bibnamefont {Held}},\ }\href {\doibase 10.1103/PhysRevResearch.3.013149} {\bibfield  {journal} {\bibinfo  {journal} {Phys. Rev. Research}\ }\textbf {\bibinfo {volume} {3}},\ \bibinfo {pages} {013149} (\bibinfo {year} {2021})}\BibitemShut {NoStop}%
\bibitem [{\citenamefont {Rubtsov}\ \emph {et~al.}(2005)\citenamefont {Rubtsov}, \citenamefont {Savkin},\ and\ \citenamefont {Lichtenstein}}]{rubtsov05}%
  \BibitemOpen
  \bibfield  {author} {\bibinfo {author} {\bibfnamefont {A.~N.}\ \bibnamefont {Rubtsov}}, \bibinfo {author} {\bibfnamefont {V.~V.}\ \bibnamefont {Savkin}}, \ and\ \bibinfo {author} {\bibfnamefont {A.~I.}\ \bibnamefont {Lichtenstein}},\ }\href {\doibase 10.1103/PhysRevB.72.035122} {\bibfield  {journal} {\bibinfo  {journal} {Phys. Rev. B}\ }\textbf {\bibinfo {volume} {72}},\ \bibinfo {pages} {035122} (\bibinfo {year} {2005})}\BibitemShut {NoStop}%
\bibitem [{\citenamefont {Werner}\ \emph {et~al.}(2006)\citenamefont {Werner}, \citenamefont {Comanac}, \citenamefont {de' Medici}, \citenamefont {Troyer},\ and\ \citenamefont {Millis}}]{werner06}%
  \BibitemOpen
  \bibfield  {author} {\bibinfo {author} {\bibfnamefont {P.}~\bibnamefont {Werner}}, \bibinfo {author} {\bibfnamefont {A.}~\bibnamefont {Comanac}}, \bibinfo {author} {\bibfnamefont {L.}~\bibnamefont {de' Medici}}, \bibinfo {author} {\bibfnamefont {M.}~\bibnamefont {Troyer}}, \ and\ \bibinfo {author} {\bibfnamefont {A.~J.}\ \bibnamefont {Millis}},\ }\href {\doibase 10.1103/PhysRevLett.97.076405} {\bibfield  {journal} {\bibinfo  {journal} {Phys. Rev. Lett.}\ }\textbf {\bibinfo {volume} {97}},\ \bibinfo {pages} {076405} (\bibinfo {year} {2006})}\BibitemShut {NoStop}%
\bibitem [{\citenamefont {Gull}\ \emph {et~al.}(2008)\citenamefont {Gull}, \citenamefont {Werner}, \citenamefont {Parcollet},\ and\ \citenamefont {Troyer}}]{gull08}%
  \BibitemOpen
  \bibfield  {author} {\bibinfo {author} {\bibfnamefont {E.}~\bibnamefont {Gull}}, \bibinfo {author} {\bibfnamefont {P.}~\bibnamefont {Werner}}, \bibinfo {author} {\bibfnamefont {O.}~\bibnamefont {Parcollet}}, \ and\ \bibinfo {author} {\bibfnamefont {M.}~\bibnamefont {Troyer}},\ }\href {\doibase 10.1209/0295-5075/82/57003} {\bibfield  {journal} {\bibinfo  {journal} {Europhysics Letters}\ }\textbf {\bibinfo {volume} {82}},\ \bibinfo {pages} {57003} (\bibinfo {year} {2008})}\BibitemShut {NoStop}%
\bibitem [{\citenamefont {Gull}\ \emph {et~al.}(2011)\citenamefont {Gull}, \citenamefont {Millis}, \citenamefont {Lichtenstein}, \citenamefont {Rubtsov}, \citenamefont {Troyer},\ and\ \citenamefont {Werner}}]{gull11}%
  \BibitemOpen
  \bibfield  {author} {\bibinfo {author} {\bibfnamefont {E.}~\bibnamefont {Gull}}, \bibinfo {author} {\bibfnamefont {A.~J.}\ \bibnamefont {Millis}}, \bibinfo {author} {\bibfnamefont {A.~I.}\ \bibnamefont {Lichtenstein}}, \bibinfo {author} {\bibfnamefont {A.~N.}\ \bibnamefont {Rubtsov}}, \bibinfo {author} {\bibfnamefont {M.}~\bibnamefont {Troyer}}, \ and\ \bibinfo {author} {\bibfnamefont {P.}~\bibnamefont {Werner}},\ }\href {\doibase 10.1103/RevModPhys.83.349} {\bibfield  {journal} {\bibinfo  {journal} {Rev. Mod. Phys.}\ }\textbf {\bibinfo {volume} {83}},\ \bibinfo {pages} {349} (\bibinfo {year} {2011})}\BibitemShut {NoStop}%
\bibitem [{\citenamefont {Kitatani}\ \emph {et~al.}(2019)\citenamefont {Kitatani}, \citenamefont {Sch\"afer}, \citenamefont {Aoki},\ and\ \citenamefont {Held}}]{Kitatani19}%
  \BibitemOpen
  \bibfield  {author} {\bibinfo {author} {\bibfnamefont {M.}~\bibnamefont {Kitatani}}, \bibinfo {author} {\bibfnamefont {T.}~\bibnamefont {Sch\"afer}}, \bibinfo {author} {\bibfnamefont {H.}~\bibnamefont {Aoki}}, \ and\ \bibinfo {author} {\bibfnamefont {K.}~\bibnamefont {Held}},\ }\href {\doibase 10.1103/PhysRevB.99.041115} {\bibfield  {journal} {\bibinfo  {journal} {Phys. Rev. B}\ }\textbf {\bibinfo {volume} {99}},\ \bibinfo {pages} {041115(R)} (\bibinfo {year} {2019})}\BibitemShut {NoStop}%
\bibitem [{\citenamefont {Harkov}\ \emph {et~al.}(2021{\natexlab{b}})\citenamefont {Harkov}, \citenamefont {Lichtenstein},\ and\ \citenamefont {Krien}}]{Harkov21}%
  \BibitemOpen
  \bibfield  {author} {\bibinfo {author} {\bibfnamefont {V.}~\bibnamefont {Harkov}}, \bibinfo {author} {\bibfnamefont {A.~I.}\ \bibnamefont {Lichtenstein}}, \ and\ \bibinfo {author} {\bibfnamefont {F.}~\bibnamefont {Krien}},\ }\href {\doibase 10.1103/PhysRevB.104.125141} {\bibfield  {journal} {\bibinfo  {journal} {Phys. Rev. B}\ }\textbf {\bibinfo {volume} {104}},\ \bibinfo {pages} {125141} (\bibinfo {year} {2021}{\natexlab{b}})}\BibitemShut {NoStop}%
\end{thebibliography}%

\end{document}